\documentclass[twoside]{dis09}
\usepackage[latin1]{inputenc}
\usepackage[dvips]{graphicx,epsfig,color}
\usepackage{wrapfig,rotating}
\usepackage{amssymb,amsmath,array}

\newcommand{\sub}[1]{\ensuremath{_{\mathrm{#1}}}}
\newcommand{\super}[1]{\ensuremath{^{\mathrm{#1}}}}

\newcommand{\GeV}{\ensuremath{\mathrm{GeV}}}

\newcommand{\der}{\ensuremath{\mathrm{d}}}
\newcommand{\pom}{\ensuremath{\Bbb{P}}}

\newcommand{\mpipi}{\ensuremath{m_{\mathrm{\pi\pi}}}}

\newcommand{\ZEUSAlphaZero}{1.096 \pm 0.021}
\newcommand{\ZEUSAlphaPrime}{0.125 \pm 0.038}

\newcommand{\OurAlphaZero}{1.093 \pm 0.003 (stat.) ^{+0.008}_{-0.007} (syst.)}
\newcommand{\OurAlphaPrime}{0.116 \pm 0.027 (stat.) ^{+0.036}_{-0.046} (syst.)}

\newcommand{\GlobalAlphaZero}{1.0871 \pm 0.0026 (stat.) \pm 0.0030 (syst.)}
\newcommand{\GlobalAlphaPrime}{0.126 \pm 0.013 (stat.) \pm 0.012 (syst.)}

\begin{document}
\title{ \boldmath
    Extraction of the Pomeron Trajectory from a Global Fit 
    to Exclusive $\rho^0$ Meson Photoproduction Data}

\author{Benno List%
%
\thanks{Supported by the German Federal Ministry of Science and
Technology under grant 05H16GUA.} ~for the H1 Collaboration
%
\vspace{.3cm}\\
%
University of Hamburg - Institute for Experimental Physics \\
Luruper Chaussee 149, D--22603 Hamburg - Germany
}

\maketitle

\begin{abstract}
Based on data on elastic $\rho^0$ photoproduction
from the H1, Omega and ZEUS collaborations,
a fit has been performed to extract the value 
$\alpha\sub\pom\,(t)$ of the pomeron trajectory
at fixed values of $t$ from the $W$ dependence of the 
differential $\gamma p$ cross section
$\der \sigma_{\gamma p}\,(W) / \der t$.
The data used in  the fit cover the range of $8.3 \le W \le 94\,\GeV$ in $\gamma p$ 
centre-of-mass energy and $0.01 \le |t| \le 0.95\,\GeV^2$ in
momentum transfer.

A linear fit to the resulting values of $\alpha\sub\pom\,(t)$
yields
$\alpha\sub{\pom,0} =\GlobalAlphaZero$ for the intercept
and $\alpha'\sub{\pom} =\GlobalAlphaPrime\,\GeV^{-2}$
for the slope of the pomeron trajectory.
The data are also compatible with the Donnachie-Landshoff 
trajectory $\alpha\sub\pom\,(t) = 1.0808 + 0.25\,\GeV^{-2}\cdot t$
at low values $|t| \lesssim 0.3\,\GeV^2$ and with a constant value
of $\alpha\sub\pom\,(t)$ at larger values of $|t|$.
\end{abstract}

\section{Introduction}

The exclusive photoproduction of $\rho^0$ mesons 
$\gamma p \to \rho^0 p$ 
has 
been studied in great detail over the last 40 years;
this process shows the typical characteristics of
a soft diffractive reaction, i.e. a weak increase
of the cross section with the photon-proton centre-of-mass 
energy $W$ and an exponential decrease as function
of the modulus of the squared momentum transfer $|t|$.

At sufficiently high centre-of-mass energies
$W$,
elastic hadron-hadron scattering is well described
by the exchange of a single Regge trajectory,
the Pomeron,
while at lower values of $W$ 
the exchange of additional meson trajectories
becomes important.
In this high $W$ region, 
the energy dependence of elastic 
$\rho^0$ photoproduction at fixed momentum transfer $t$
is in diffractive models directly linked to the Pomeron trajectory 
$\alpha\sub\pom\,(t)$
by
$$
   \frac{\der \sigma\sub{\gamma p}\,(W)}{\der t} 
   \propto 
   \left ( \frac{W}{W\sub{0}} \right ) ^{4 (\alpha\sub\pom\,(t) - 1)}.
$$
Thus, a measurement of the $W$ dependence 
of elastic $\rho^0$ photoproduction in bins of $t$
determines the 
Pomeron trajectory $\alpha\sub\pom\,(t)$,
which is often approximated by a linear function
$\alpha\sub\pom\,(t) = \alpha\sub{\pom, 0} + \alpha'\sub\pom \cdot t$.

In the present analysis \cite{url,bib:H1prelim-09-016},
we combine in a global fit
the new H1 measurements \cite{bib:h1-05}
with earlier measurements of the H1 \cite{Aid:1996bs}, Omega \cite{Aston:1982hr}
and the ZEUS collaborations \cite{Derrick:1996vw, Breitweg:1997ed, Breitweg:1999jy}
in order to achieve the best possible accuracy
for the determination of the Pomeron trajectory.
By including data from other experiments, we can improve the accuracy
compared to a fit to the H1 data alone,
which was performed earlier \cite{bib:h1-05}.
Such a combination was first performed by the ZEUS collaboration \cite{Breitweg:1999jy}.

\section{Input data sets}

\subsection{The Omega data set}

In 1982, the Omega collaboration published \cite {Aston:1982hr} 
measurements of elastic $\rho^0$ photoproduction 
at a tagged photon beam at CERN.
The measurements were performed in the kinematic range $0.06 < |t| < 1\,\GeV^2$
in three bins of photon energy $E_\gamma$, 
corresponding to $\gamma p$ centre-of-mass energies of
$6.8$, $8.3$, and $10.3\,\GeV$, respectively.
The measurements were restricted to the di-pion mass range
$0.56 < m_{\pi\pi} < 0.92\,\GeV$.

Differential cross sections were measured in $47$ bins of $t$.
Unfortunately, no table of 
the original cross sections with their errors was  
published,
only the result of the fits.
In order to include the data into the fit,
the original event numbers were extracted from Figure 3 of the paper.
The fit of the $t$ dependence given in the paper was repeated, with identical results;
however, the statistical errors had to be increased by a 
factor of
$\sqrt{\chi^2/n.d.f.}$ in order to obtain reasonable $\chi^2$ values of 
these fits. 

The factor between number of events and the differential cross section per bin
was determined from the published $t$-integrated cross sections, 
such that the integral of the fitted differential cross section
in the range $0 < |t| < 1\,\GeV^2$ reproduces the published integrated cross section
as obtained from a fit using the Ross-Stodolsky line shape
\cite {Ross:1965qa}.

An additional correction factor 
of $1.187 \pm 0.051$ was applied in order to correct the cross section
to the full di-pion mass range of $2 m\sub{\pi} < \mpipi < 1.52\,\GeV$,
which is the cross section definition adopted in this analysis.
The error is due to a variation of the skewing
parameter $n$ in the region $1 < n < 4$, and
is treated as an  uncorrelated systematic error.

Finally, the data points were grouped according to the $t$ binning used for the 
global fit, corrected to the bin centre in $t$ according to the $t$ dependence 
given by the Omega collaboration, and averaged.

A global normalization error, which is assumed to be fully correlated 
for all data points in a given photon energy interval,
is derived from the uncertainty of the integrated cross section
as given by the Omega collaboration.

\subsection{The H1 data sets}

\subsubsection*{The H1 HERA-1 measurement}

The H1 Collaboration has measured elastic $\rho^0$ photoproduction 
\cite{Aid:1996bs}
for $0 < |t| < 0.5\,\GeV^2$
at a mean $\langle W \rangle = 55\,\GeV$ using data taken in 1993.
Of the systematic errors listed in Tab.~4 of the paper,
the errors from the track fit efficiency, resonance extraction and
luminosity as well as half of the error from the trigger efficiency were
taken as normalization uncertainty of $17\,\%$,
while the rest of the errors are treated as uncorrelated, amounting to
$22\,\%$.

The data points were corrected for the $t$ slope according to the
measured $t$ dependence and averaged where appropriate, treating the
systematic error of the averaged data points as fully correlated.

\subsubsection*{The preliminary H1 HERA-2 measurement}

Based on data taken in 2005, the H1 collaboration has released
preliminary data \cite{bib:h1-05}, covering the range of $W=20 -
90\,\GeV$ and $0 <|t|<0.7\,\GeV^2$ for the measurement of
the elastic cross section; 60 cross section values have been measured
in eight bins of $t$, with ten  bins in $W$ for
$|t|<0.1\,\GeV^2$ and five at larger values of $|t|$.
Ten sources of correlated errors, as described in \cite{bib:h1-05},
are taken into account.
The normalization uncertainty of this data set amounts to $5.3\,\%$.

These data are sufficient to determine $\alpha\sub{\pom}\,(t)$ with
high accuracy for $|t|<0.4\,\GeV^2$.

\subsection{The ZEUS data sets}

\subsubsection*{The ZEUS LPS measurement}

The ZEUS collaboration has used their Leading Proton Spectrometer (LPS)
in a measurement of elastic $\rho$ photoproduction \cite{Derrick:1996vw}
using data taken in 1994.
Four data points with $t$ values in the range
$0.073 < |t| < 0.40\,\GeV^2$ were published.
The measurement covered a range in photon-proton centre-of-mass energy of
$50 < W < 100\,\GeV$, with a mean $\langle W \rangle = 73\,\GeV$.

From the list of systematic uncertainties published, we treat the errors from
the luminosity, sensitivity to the proton beam angle, cross section extraction,
radiative corrections and background from $\omega$ and $\phi$ production
as fully correlated, corresponding to an overall normalization error of
$6\,\%$; the rest of the errors is treated as an uncorrelated error
of $11\,\%$.

The published differential cross section values were corrected to the 
nearest $t$ bin
centre according to the $t$-dependence observed by ZEUS.
The error from this {\em swimming} of the data points was evaluated and is 
treated as a correlated systematic error.

The data points were also corrected from the mass range $0.55 < m_{\pi\pi} < 1.2\,\GeV$
used in the publication to the full mass range as described above; 
the correction factor amounts to $1.09$.
Since the published data are already corrected for skewing effects, 
no additional 
uncertainty arises
here.

\subsubsection*{The ZEUS low-$|t|$ measurement}

Using data taken from the year 1994, the ZEUS collaboration
has performed a measurement of elastic $\rho$ photoproduction 
\cite{Breitweg:1997ed} in the region $|t|<0.5\,\GeV$ and
$50 < W < 100\,\GeV$, at a mean value of $\langle W \rangle = 71.7\,\GeV$.
Differential cross sections were measured in twelve $t$ bins.

We have averaged the first three $t$ bins and the fourth and fifth
$t$ bin and 
corrected the measurements to the $t$ binning chosen for this global fit,
using the published $t$ dependence.

The systematic errors from the luminosity measurement, the cross section extraction and 
the radiative corrections were assumed to be correlated,
resulting in a $5\,\%$ normalization uncertainty,
the rest of the systematic errors are treated as uncorrelated and amount to $10\,\%$.

\subsubsection*{The ZEUS high-$|t|$ measurement}

Based on a data set taken in 1995,
the ZEUS collaboration has performed a third measurement
of elastic $\rho^0$ photoproduction \cite{Breitweg:1999jy},
covering the range $0.35 < |t| < 1.62\,\GeV^2$.

The seven data points in the range $0.35 < |t| < 1.0\,\GeV^2$ were 
transferred to the common $t$ bin centres of this global analysis if neccessary,
using the published $t$ dependence.

A global normalization uncertainty of $15\,\%$ and the published values
for the other systematic errors were taken into account;
the uncertainty arising from the subtraction of proton dissociation 
background was considered as a correlated error.

\section{Fit Procedure}

\begin{figure}[b!]
\center
\setlength{\unitlength}{1cm}
\epsfig{file=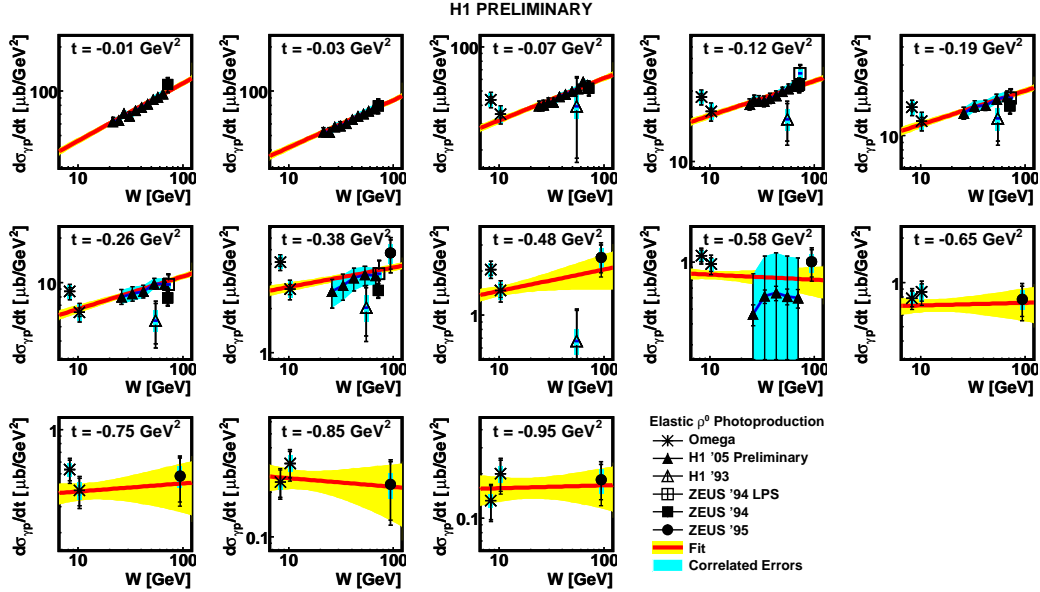,width=14cm}
\caption[]{The fit of the form 
$\der \sigma / \der t (\gamma p \to \rho^0 p) = s\sub i (W/W_0)^{4 (\alpha\sub i - 1)}$
with $W_0 = 40\,\GeV$ to the elastic $\rho^0$ photoproduction 
cross section data
from the H1 \cite{bib:h1-05, Aid:1996bs}, Omega \cite{Aston:1982hr}
and ZEUS \cite {Derrick:1996vw, Breitweg:1997ed, Breitweg:1999jy}
collaborations.
For the data points, the inner error bars represent the statistical and uncorrelated
systematic error, the outer error bars the full error; the shaded band indicates 
the size of the correlated error of each data set.

The global fit has a $\chi^2/d.f. = 111.7 / 80$.
}
\label{fig:1}
\end{figure}

The differential $\gamma p$ cross sections 
measured at a particular
value of 
$t = t\sub j$ are fitted to a function 
\begin{equation}
  \label{eq:wdep}
  \frac{\der \sigma\super{\gamma p}\,(W)}{\der t}
  = s\sub j
    \left ( \frac {W\sub i}{W_0} \right ) ^{4 (\alpha\sub j - 1)} = f\sub i,
\end{equation}
with 
$W_0 = 40\,\GeV$.
The parameter $\alpha\sub j$ is, for elastic $\rho^0$ production,
the value of the pomeron trajectory $\alpha\sub j = \alpha\sub\pom\,(t\sub j)$,
and $s\sub j = \frac{\der \sigma\super{\gamma p}\,(W_0, t\sub j)}{\der t}$
is the differential cross section at the respective value of $t$.

The fit minimizes a $\chi^2$ expression that takes into account correlated systematic
errors and
 is similar to the method employed in the averaging of
inclusive $F_2$ data \cite{Aaron:2009bp}. 

Altogether $26$ parameters $a\sub j$ are determined, corresponding to the normalization and the $W$
exponent in each of the $13$ $t$ bins,
plus $19$ coefficients $b\sub k$ for the correlated error sources.

It is important to note that the global fit is not based on any assumption
about the $t$ dependence, i.e. the form factor, of the cross section,
but that for each $t$ bin the normalization $s\sub i$ and the $W$ 
exponent $\alpha\sub i$
are determined individually.

However, because the fit includes all data points together,
correlated error sources lead to 
correlations between the resulting fit parameters
across different $t$ bins.

Correlated and uncorrelated errors are fully propagated in this fit,
so that the full covariance matrix is obtained 
for the resulting values $\alpha\sub\pom\,(t\sub i)$.

Using the values  $\alpha\sub\pom\,(t\sub i)$ obtained from the fit to
the data and their covariance matrix, it is possible to 
perform a straight line fit to obtain the intercept $\alpha\sub{\pom,0}$
and slope $\alpha'\sub{\pom}$ of the pomeron trajectory.
The $\chi^2$ of this fit may serve as an indication whether such a
straight line is able to describe the data in a satisfactory way.

\section{Results}

Fig.~\ref{fig:1} shows the result of the global fit in all 13 $t$ bins.
The overall quality of the fit is satisfactory, with a $\chi^2 =
111.7$ for $d.f.=80$ degrees of freedom (106 data points enter the fit, and 26 free parameters are
determined). 

All data sets contribute a reasonable amount to the total
$\chi^2$. 
However, it is clearly visible that the Omega data at $W=8.3\,\GeV$
lie systematically above the fit for $|t| < 0.6\,\GeV^2$, which leads to
a best value for the normalization of this data set that is
significantly shifted; the shift is, however, less than $3$ standard
deviations.

Since correlated uncertainties, in particular the normalizations, of all
data sets are treated consistently across the whole $t$ range,
the high precision of the H1 data from 2005 at low $|t|$ constrains 
the normalization of all data sets,
which reduces the uncertainty from the correlated errors
also at larger values of $|t|$.

\begin{figure}[b!]
\center
\setlength{\unitlength}{1cm}
\epsfig{file=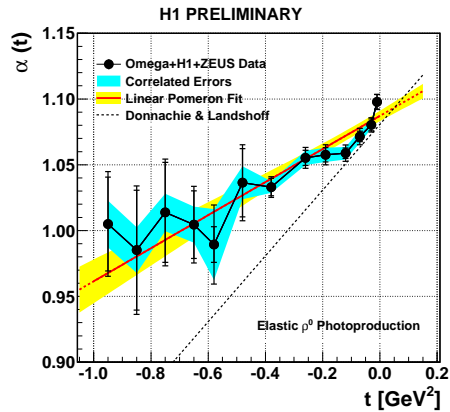,height=6.0cm}
\caption[]{The result of the fit shown in Fig.~\ref{fig:1}.
The coefficients $\alpha\sub i = \alpha\sub{\pom}\,(t\sub i)$
are shown as a function of the momentum transfer $t$.
The inner error bars represent the statistical and uncorrelated
systematic error, the outer error bars the full error; the shaded band indicates 
the size of the correlated error of each data set.

The Donnachie and Landshoff pomeron trajectory \cite{Donnachie:1992ny,bib:dlslope}
$\alpha\sub{\pom}\,(t) = 1.0808 + 0.25 \cdot t$
is  shown as dashed line.
A linear fit to the data is indicated by the straight line and shaded band.
}
\label{fig:2}
\end{figure}

Fig.~\ref{fig:2} shows the fitted values $\alpha\sub{\pom}\,(t)$
as a function of $t$.
The values are in excellent agreement with the values 
obtained from the H1 data alone \cite{bib:h1-05}
and with those obtained by the ZEUS collaboration
from a combined fit to the Omega, ZEUS, and '93 H1 data \cite{Breitweg:1999jy}.

However, while the values obtained by ZEUS are in perfect agreement 
with a linear pomeron trajectory of the form
$\alpha\sub\pom\,(t) = \alpha\sub{\pom,0} + \alpha\sub{\pom}' \cdot t$,
which has a significantly shallower slope $\alpha\sub{\pom}' = 0.125\,\GeV^{-2}$
than the canonical value \cite{bib:dlslope} $\alpha\sub{\pom}' = 0.25\,\GeV^{-2}$
 for the soft pomeron,
the new, more precise result shows that
at low values of $|t|\lesssim 0.3\,\GeV^2$ the data are in 
very good agreement with a slope of $0.25\,\GeV^{-2}$.
At larger values $|t|\gtrsim 0.5\,\GeV^2$, a significant difference between
the data and the Donnachie-Landshoff soft pomeron trajectory
is found; the data are even compatible with a constant
value of $\alpha\sub{\pom} \approx 1.00 \pm 0.03$ in this region.

These observations are in broad agreement with the indications from UA8 data \cite{bib:ua8}
that the pomeron trajectory flattens at large $|t|$.
It is interesting to note that also models based on gauge--string duality \cite{Brower:2006ea}
predict such a behaviour.

A straight line fit to the observed $\alpha\sub{\pom}$ values yields
for the Pomeron trajectory
$$
  \alpha\sub\pom\,(t) = \GlobalAlphaZero + (\GlobalAlphaPrime)\,\GeV^{-2} \cdot t
$$
with a correlation coefficient of $0.37$ between 
$\alpha\sub{\pom,0}$ and $\alpha'\sub\pom$.
This result is in excellent agreement with the
measurement from the ZEUS collaboration \cite{Breitweg:1999jy}
$
  \alpha\sub\pom\,(t) = \ZEUSAlphaZero + (\ZEUSAlphaPrime)\,\GeV^{-2} \cdot t,
$
and the measurement using H1 data alone \cite{bib:h1-05}
$
  \alpha\sub\pom\,(t) = \OurAlphaZero + (\OurAlphaPrime)\,\GeV^{-2} \cdot t.
$
The linear fit has a reasonable overall $\chi^2/d.f. = 14.7/11$,
therefore the hypothesis of a linear
pomeron trajectory in the range $0 < |t| < 1\,\GeV^2$
cannot be excluded.

\begin{footnotesize}


\end{footnotesize}



\begin{thebibliography}{99}

\bibitem{url} Slides: \\ 
\verb$http://indico.cern.ch/contributionDisplay.py?contribId=161&sessionId=18&confId=53294$

\bibitem{bib:H1prelim-09-016}
  H1 Collaboration,
  H1prelim-09-016 (2009),\\
\verb$http://www-h1.desy.de/h1/www/publications/htmlsplit/H1prelim-09-016.long.html$

\bibitem{bib:h1-05}
  H1 Collaboration,
  H1prelim-06-011 (2006),\\
\verb$http://www-h1.desy.de/publications/htmlsplit/H1prelim-06-011.long.html$\\
%
  R.~M.~Weber,
  ``Diffractive $\rho^0$-photoproduction at H1,''
  PhD thesis ETH Zurich 2006, DISS-ETH-16709.

\bibitem{Aid:1996bs}
  S.~Aid {\it et al.}  [H1 Collaboration],
  Nucl.\ Phys.\  {\bf B463} 3 (1996).
  
\bibitem{Aston:1982hr}
  D.~Aston {\it et al.},
  Nucl.\ Phys.\  {\bf B209} 56 (1982).

\bibitem{Derrick:1996vw}
  M.~Derrick {\it et al.}  [ZEUS Collaboration],
  Z.\ Phys.\  {\bf C73} 253 (1997).

\bibitem{Breitweg:1997ed}
  J.~Breitweg {\it et al.}  [ZEUS Collaboration],
  Eur.\ Phys.\ J.\  {\bf C2} 247 (1998).

\bibitem{Breitweg:1999jy}
  J.~Breitweg {\it et al.}  [ZEUS Collaboration],
  Eur.\ Phys.\ J.\  {\bf C14} 213 (2000).
      
\bibitem{Ross:1965qa}
  M.~Ross and L.~Stodolsky,
  Phys.\ Rev.\  {\bf 149} 1172 (1966).

\bibitem{Aaron:2009bp}
  F.~D.~Aaron {\it et al.} [H1~Collaboration],
  arXiv:0904.0929 [hep-ex] (2009).
  
\bibitem{Donnachie:1992ny}
  A.~Donnachie and P.~V.~Landshoff,
  Phys.\ Lett.\  {\bf B296} 227 (1992).

\bibitem{bib:dlslope}
  G.~A.~Jaroszkiewicz and P.~V.~Landshoff,
  Phys.\ Rev.\  {\bf D10} 170 (1974);
\\
  P.~V.~Landshoff,
  Nucl.\ Phys.\ Proc.\ Suppl.\  {\bf 12} 397 (1990).


\bibitem{bib:ua8}
  A.~Brandt {\it et al.}  [UA8 Collaboration],
  Nucl.\ Phys.\   {\bf B514} 3 (1998);\\
%
  S.~Erhan and P.~E.~Schlein,
  Phys.\ Lett.\   {\bf B481} 177 (2000).
  
\bibitem{Brower:2006ea}
  R.~C.~Brower, J.~Polchinski, M.~J.~Strassler and C.~I.~Tan,
  JHEP {\bf 0712} 005 (2007).
  

\end{thebibliography}
\end{document}